# Distribution fitting 13. Analysis of independent, multiplicative effect of factors. Application to effect of essential oils extracts from plant species on bacterial species. Application to the factors of antibacterial activity of plant species

**Lorentz JÄNTSCHI**[1,3,*], **Sorana D. BOLBOACĂ**[2], **Mugur C. BĂLAN**[3], **Radu E. SESTRAŞ**[1]

[1] University of Agricultural Sciences and Veterinary Medicine Cluj-Napoca
[2] "Iuliu Haţieganu" University of Medicine and Pharmacy Cluj-Napoca
[3] Technical University of Cluj-Napoca
[*] Corresponding author: lori@academicdirect.org

**Abstract.** A factor effect study was conducted on a set of observations at the contingency of a series of plant species and bacteria species regarding the antibacterial activity of essential oil extracts. The study reveals a very good agreement between the observations and the hypothesis of independent and multiplicative effect of plant and bacteria species factors on the antibacterial activity. Shaping of the observable to a Negative Binomial distribution allowed the separation of two convoluted Gamma distributions in the observable further assigned to the distribution of factors. Statistics of the Gamma distribution allowed estimating the ratio between diversity of plants factors and bacteria factors in the antibacterial activity of essential oils extracts.

## Introduction

Recoding the data from an observation may provide different types of outcomes: binary, multinomial, or ordinal, if are seen different states of the observed; absolute or relative values if a measurement scale or ratio are used; more, the data may come from a discrete or continuous pool of possible values and our observation may or may not catch the true or whole domain of the observable and some times we can miss even its type. This is the main reason for which assumptions are made and statistics are involved to check the assumptions at a certain level of confidence.

Going forward with the observation, experiments are designed in order to collect the data in certain imposed conditions allowing us to extract the information regarding the observed variable or phenomena.

Agreement between a model and a series of observations usually implies estimation of unknown (unforeseen) parameters about which we may assume that are characteristics of the population of whole possible observations from which the sample of observations were drawn (Jäntschi, 2009[1]). Measuring the agreement between the model and the series of observation is a matter of statistics, requires a given specific model and a series of statistical tests, designed for general or specific cases give different measures of the agreement, based too on other certain assumptions regarding the observed phenomena (Jäntschi & Bolboacă, 2009[2]). The most common assumption is the assumption of normality and it comes from the common sense that most of the data that we observe are normal distributed, or it comes from populations that are normal distributed. But even in this case a global agreement of all statistics involved occurs far less than our expectations and should apply when a conclusion regarding the agreement between the observation and the model are drawn (Bolboacă & Jäntschi, 2009[3]).

Not always the measurement and the analysis are conducted by the same people or same group of people. In certain cases, the data may suffer alteration processes during the way from observation or experiment to analysis. Certain statistics were developed to cover this aspect too, and to measure the probability that data may not come from an impartial observation (Jäntschi & others, 2009a[4]).

Certain conditions imposed to the experiment collecting the observation may reshape the original distribution of the observed population, and by using distribution analysis is possible to obtain this new shape, specifically to the experiment in which the observation were made (Jäntschi & others, 2009h[5]). In other cases, subject of observation may provide unsymmetrical shape of observed values, giving weight to higher (or lower) values disfavouring the opposite case and this fact can be revealed too (Marta & others, 2009[6]).

Contingency of factors in the observable is one of the most important aspects of experimental studies, and may indicate or proof the way in which a process should be conducted in order to maximize the outcome (Jäntschi & Bălan, 2009[7]). Field experiments are usually conducted in environmental conditions which are not in control of the observer, and knowing the inferences coming from changing of these conditions is essential to the parsimony of the factors affecting the observable (Bălan & others, 2009[8]).

Other aspects such as special cases in which values recorded in an ordered outcome category based experiment may provide useful knowledge about the corrections that should be made on the values associated with the categories (Stoenoiu & others, 2009[9]; Stoenoiu & others, 2010[10]).

Analysing the data regarding the morphology of plants spread in a certain region provide knowledge about the effect of adaptation to certain conditions of living plant species (Jäntschi & Bolboacă, 2011[11]). Other effects of environmental conditions in which plants forced to be adapted can be found from distribution of chemical compounds in plant species (Jäntschi & others, 2011[12]).

The present study takes into the analysis the distribution of the antibacterial activity at the contingence between plant species and essential oil extracts of plant species. The aim of the study is to reveal how the biological activity is influenced by plant and bacteria species and to infer the distribution of biological activity from plant and bacteria species.

**Material**

The data regarding the antibacterial activity of essential oils of plants measured as inhibition zones by using disc-diffusion method are taken from an experimental study (Soković & others, 2007[13]) and are given in Table 1.

Tab.1

Antibacterial activity of essential oils plant extracts on bacteria

| Antibacterial activity - inhibition zones in mm | | Plant species | | | | | | | | | |
|---|---|---|---|---|---|---|---|---|---|---|---|
| | | M.s. | M.p. | C.l. | C.a. | M.c. | L.a. | O.b. | S.o. | O.v. | T.v. |
| Bacteria species | M.flavus | 25 | 25 | 19 | 19 | 13 | 22 | 23 | 15 | 35 | 30 |
| | B.subtilis | 24 | 22 | 18 | 18 | 12 | 20 | 22 | 14 | 34 | 28 |
| | S.epidermidis | 20 | 20 | 14 | 14 | 12 | 18 | 18 | 12 | 30 | 26 |
| | S.aureus | 22 | 20 | 16 | 14 | 10 | 18 | 18 | 12 | 32 | 28 |
| | S.enteritidis | 20 | 20 | 13 | 10 | 9 | 16 | 18 | 10 | 27 | 24 |
| | S.typhimurium | 18 | 17 | 11 | 8 | 8 | 16 | 16 | 10 | 25 | 20 |
| | E.coli | 16 | 16 | 12 | 9 | 9 | 14 | 14 | 10 | 26 | 22 |
| | E.cloacae | 14 | 14 | 9 | 9 | 9 | 12 | 12 | 10 | 25 | 22 |
| | L.monocytogenes | 16 | 13 | 9 | 8 | 8 | 10 | 11 | 9 | 25 | 18 |
| | P. mirabilis | 10 | 11 | 0 | 0 | 0 | 7 | 8 | 0 | 22 | 18 |
| | P. aeruginosa | 10 | 10 | 0 | 0 | 0 | 6 | 8 | 0 | 20 | 16 |
| M.s.: Mentha spicata; M.p.: Mentha piperita; C.l.: Citrus limon; C.a.: Citrus aurantium; M.c.: Matricaria chamommilla; L.a.: Lavandula angustifolia; O.b.: Ocimum basilicum; S.o.: Salvia officinalis; O.v.: Origanum vulgare; T.v.: Thymus vulgaris; | | | | | | | | | | | |

On the data from Table 1 an analysis of independence were conducted using Chi-square test. The analysis revealed that for P. mirabilis and P. aeruginosa bacteria species the hypothesis of independence cannot be accepted (d.f.$_{Plants}$ = 9; $X^2_{P.\ mirabilis}$ = 29.1; $X^2_{P.\ aeruginosa}$ = 26.2; $p_{\chi 2}$(P. mirabilis) < 1‰; $p_{\chi 2}$(P. aeruginosa) < 2‰) and therefore were withdrawn from further analysis.

Without these two bacteria, the analysis of independence was conducted again, when the $X^2$ statistic decreased dramatically ($X^2$(10-1 plants, 11-1 bacteria) = 69.3; $X^2$(10-1 plants, 9-1 bacteria) = 8.5; $p_{\chi 2}$(8.5,72) > 0.9999).

As can be observed, the data given in Table 1 are integers (millimetres) and then even if the true distribution of the observable is not a discrete one, the observed distribution is always discrete when we use an instrumentation that has a precision limit of one millimetre.

For simplicity (an in the mean time for generality), let's note with `m` the number of rows - bacteria (m = 9) and with `n` the number of cols - plants (n = 10). Let us recall that the expectances under assumption of independence between rows and cols are given by (where $O_{i,j}$ are the observed cell from $i^{th}$ row and $j^{th}$ column in Table 1:

$$E_{i,j} = \sum_{k=1}^{m} O_{i,k} \sum_{k=1}^{n} O_{k,j} \Big/ \sum_{i=1}^{m} \sum_{j=1}^{n} O_{i,j}$$

**Method**

Analysis of multiplicative effect of factors under assumption of normal distributed observed absolute error (Fisher, 1923[14]; Bolboacă & others, 2011[15]) give the following equation relating the "$a_i$" - rows factors and "$b_j$" - cols factors:

$$E_{i,j} = a_i \cdot b_j; \ S^2 = \sum_{i=1}^{m} \sum_{j=1}^{n} (O_{i,j} - E_{i,j})^2 = \sum_{i=1}^{m} \sum_{j=1}^{n} (O_{i,j} - a_i \cdot b_j)^2 = \min.$$

This equation leads very easy (derivatives should be null in the minimum) to a system of equations. Unfortunately, its major disadvantage is that admits infinity (a simple infinity) of solutions (i.e. for any fixed $a_1$ has only one solution). Its minor disadvantage is that trying to express all other variables depending on one of them (or all depending to a parameter) leads to polynomials of degree min(m,n) without simple form in the general case. Thus one way in which the solution may be found (only in numerical case) is guessing a starting value and iterating directly from the system of equations.

Since all values are relative to one of them, starting values give only one solution (the nearest one). We choose to start with $a_i^0$ given by following formula, and then iterate repeatedly with ($b_j^1$; $a_i^1$), ($b_j^2$; $a_i^2$), …, until $S^2$ converged:

$$a_i^0 = \sqrt{\sum_{i=1}^{m} \sum_{j=1}^{n} O_{i,j} \Big/ r} \ ; \left( a_i = \sum_{j=1}^{n} b_j O_{i,j} \Big/ \sum_{j=1}^{n} b_j^2, i=1..m; \ b_j = \sum_{i=1}^{m} a_i O_{i,j} \Big/ \sum_{i=1}^{m} a_i^2, j=1..n \right)$$

The solution (rows and columns factors) is given in Table 2.

**Results and Discussion**

Tab.2

Row and column factors in data from Table 1

| row(i) | 1 | 2 | 3 | 4 | 5 | 6 | 7 | 8 | 9 | |
|---|---|---|---|---|---|---|---|---|---|---|
| $a_i$ | 5.637 | 5.3042 | 4.6295 | 4.832 | 4.2522 | 3.7921 | 3.7769 | 3.4933 | 3.2801 | |
| column(j) | 1 | 2 | 3 | 4 | 5 | 6 | 7 | 8 | 9 | 10 |
| $b_j$ | 4.4859 | 4.2914 | 3.1425 | 2.8649 | 2.3001 | 3.766 | 3.9267 | 2.6094 | 6.585 | 5.5636 |

A distribution analysis can be conducted on the whole data from Table 1, which passed the independence test. By taking into account that the data are integers only, a suitable distribution is a discrete one. Table 3 contains the analysis with discrete type distribution alternatives.

Tab. 3

Distribution of the observed antibacterial activity

| Distribution | Parameters | K-S | $p_{K-S}$ | A-D | $p_{A-D}$ | C-S | $p_{C-S}$ |
|---|---|---|---|---|---|---|---|
| Uniform | a=6 b=28 | 0.13043 | 0.09 | 22.35 | $9.7 \cdot 10^{-8}$ | 19 | $9.3 \cdot 10^{-5}$ |
| Geometric | p=0.05525 | 0.40041 | $2.8 \cdot 10^{-13}$ | 17.783 | $6.7 \cdot 10^{-7}$ | 43 | $4.3 \cdot 10^{-10}$ |
| Logarithmic | θ=0.98663 | 0.60571 | 0 | 41.427 | 0 | ∞ | 0 |
| Neg. Binomial | r=11 p=0.609 | 0.08675 | 0.481 | 0.80817 | 0.408 | 1.63 | 0.443 |
| Poisson | λ=17.1 | 0.18105 | $4.7 \cdot 10^{-3}$ | 10.152 | $4.8 \cdot 10^{-5}$ | 15 | $4.8 \cdot 10^{-4}$ |
| Bernoulli, Binomial, Hypergeometric: No MLE fit; C-S=ln(1/$p_{K-S}$)+ln(1/$p_{A-D}$) | | | | | | | |

Results given in Table 3 clearly indicate that the distribution of the antibacterial activity of essential oils extracts among bacteria is of negative binomial type. A mathematical analysis of the negative binomial distribution allows explaining of this fact. Thus, a simple math gives:

$$NegBin(x;r,p) = \int_0^\infty Poisson(x;z) \cdot Gamma(z;r,\frac{p}{1-p}) dz,$$

where

$$NegBin(x;r,p) = \frac{\Gamma(r+x)p^x(1-p)^r}{\Gamma(x+1)\Gamma(r)}, \quad Gamma(z;\alpha,\beta) = \frac{z^{\alpha-1}e^{-z/\beta}}{\beta^\alpha \Gamma(\alpha)}, \quad Poisson(x;\lambda) = \frac{\lambda^x e^{-\lambda}}{\Gamma(x+1)}$$

The previous formula shows that the negative binomial distribution may arise as a continuous mixture of Poisson distributions where the mixing distribution of the Poisson rate is a gamma distribution. Thus, under these assumption that it should be behind of this distribution a mixture of Poisson and Gamma distributions, between parameters should be the previous proofed formula. Indeed, the data behave this property. Any row and any column agree with a certain Poisson distribution (Table 4).

Tab. 4

Poisson shaping of the observed series of data

| Species | Data | MLE | λ | $p_{K-S}$ | $p_{A-D}$ | C-S | $p_{C-S}$ |
|---|---|---|---|---|---|---|---|
| M.flavus | 25; 25; 19; 19; 13; 22; 23; 15; 35; 30 | -33.354 | 22.6 | 0.96235 | 0.51167 | 0.708 | 0.7017 |
| B.subtilis | 24; 22; 18; 18; 12; 20; 22; 14; 34; 28 | -33.054 | 21.2 | 0.97719 | 0.54082 | 0.638 | 0.7270 |
| S.epidermidis | 20; 20; 14; 14; 12; 18; 18; 12; 30; 26 | -31.866 | 18.4 | 0.65898 | 0.40743 | 1.315 | 0.5182 |
| S.aureus | 22; 20; 16; 14; 10; 18; 18; 12; 32; 28 | -34.557 | 19.0 | 0.84516 | 0.33102 | 1.274 | 0.5289 |
| S.enteritidis | 20; 20; 13; 10; 9; 16; 18; 10; 27; 24 | -33.470 | 16.7 | 0.51709 | 0.20678 | 2.236 | 0.3270 |
| S.typhimurium | 18; 17; 11; 8; 8; 16; 16; 10; 25; 20 | -31.897 | 14.9 | 0.37285 | 0.27307 | 2.285 | 0.3191 |
| E.coli | 16; 16; 12; 9; 9; 14; 14; 10; 26; 22 | -31.350 | 14.8 | 0.884 | 0.3888 | 1.068 | 0.5863 |
| E.cloacae | 14; 14; 9; 9; 9; 12; 12; 10; 25; 22 | -31.372 | 13.6 | 0.74308 | 0.21328 | 1.842 | 0.3981 |
| L.monocytogenes | 16; 13; 9; 8; 8; 10; 11; 9; 25; 18 | -31.259 | 12.7 | 0.63353 | 0.27459 | 1.749 | 0.4171 |
| M.s. | 25; 24; 20; 22; 20; 18; 16; 14; 16 | -24.538 | 19.444 | 0.93832 | 0.70654 | 0.411 | 0.8142 |
| M.p. | 25; 22; 20; 20; 20; 17; 16; 14; 13 | -24.633 | 18.556 | 0.59374 | 0.67567 | 0.913 | 0.6334 |
| C.l. | 19; 18; 14; 16; 13; 11; 12; 9; 9 | -23.817 | 13.444 | 0.98545 | 0.71802 | 0.346 | 0.8412 |
| C.a. | 19; 18; 14; 14; 10; 8; 9; 9; 8 | -25.167 | 12.111 | 0.69926 | 0.37978 | 1.326 | 0.5153 |
| M.c. | 13; 12; 12; 10; 9; 8; 9; 9; 8 | -20.013 | 10 | 0.21701 | 0.32018 | 2.667 | 0.2636 |
| L.a. | 22; 20; 18; 18; 16; 16; 14; 12; 10 | -24.412 | 16.222 | 0.74614 | 0.71089 | 0.634 | 0.7283 |
| O.b. | 23; 22; 18; 18; 18; 16; 14; 12; 11 | -24.965 | 16.889 | 0.69564 | 0.66245 | 0.775 | 0.6788 |
| S.o. | 15; 14; 12; 12; 10; 10; 10; 10; 9 | -20.654 | 11.333 | 0.28949 | 0.3188 | 2.383 | 0.3038 |
| O.v. | 35; 34; 30; 32; 27; 25; 26; 25; 25 | -25.626 | 28.778 | 0.41834 | 0.50869 | 1.547 | 0.4613 |
| T.v. | 30; 28; 26; 28; 24; 20; 22; 22; 18 | -25.333 | 24.222 | 0.9651 | 0.6874 | 0.410 | 0.8145 |
| C-S=ln(1/$p_{K-S}$)+ln(1/$p_{A-D}$); Σln(1/$p_{C-S}$)=12.3; $p_{C-S\text{-"Poisson"}}$ = 0.8741 | | | | | | | |

Results obtained so far show that two parts out of three results directly from the analysis of the distribution of observed data (Negative Binomial distribution of the whole pool of independent data; Poisson distribution of the series of independent data). More, let's note that with the data from Table 4, Average($\lambda$) for Bacteria is 17.1000 and Average($\lambda$) for Plants is 17.0999. It remains only that Poisson parameters of the series to be Gamma distributed. Indeed, results given in Table 5 proof this fact.

Tab.5
Maximum Likelihood Estimation (MLE) of Poisson parameters of species from Table 4

| Hypothesis | $\partial LE/\partial r= \partial LE/\partial p=0$ | r (Natural) | p($\partial LE/ \partial p=0$) | p/ (1-p) | MLE | $p_{K-S}$ | $p_{A-D}$ | $p_{C-S}$ | C-S | $p_{C-S}$ |
|---|---|---|---|---|---|---|---|---|---|---|
| $\lambda_A \sim$ Gamma(r,p/(1-p)) | r=14.127; p=0.547 | 10 | 0.631 | 1.710 | -55.801 | 0.993 | 0.833 | 0.917 | 0.276 | 0.964 |
| | | 11 | 0.609 | 1.555 | -55.561 | 0.997 | 0.878 | 0.948 | 0.187 | 0.980 |
| | | 12 | 0.588 | 1.425 | -55.401 | 0.999 | 0.902 | 0.974 | 0.130 | 0.988 |
| | | 13 | 0.568 | 1.315 | -55.310 | 0.999 | 0.909 | 0.865 | 0.241 | 0.971 |
| | | **14** | **0.550** | **1.221** | **-55.277** | **0.998** | **0.901** | **0.846** | **0.273** | **0.965** |
| | | 15 | 0.533 | 1.140 | -55.293 | 0.990 | 0.880 | 0.490 | 0.851 | 0.837 |
| $\lambda_P \sim$ Gamma(r,p/(1-p)) | r=9.788; p=0.636 | 9 | 0.655 | 1.900 | -30.843 | 0.995 | 0.853 | 0.869 | 0.304 | 0.959 |
| | | **10** | **0.631** | **1.710** | **-30.826** | **0.984** | **0.827** | **0.886** | **0.327** | **0.955** |
| | | 11 | 0.609 | 1.555 | -30.862 | 0.961 | 0.788 | 0.901 | 0.382 | 0.944 |
| | | 12 | 0.588 | 1.425 | -30.940 | 0.929 | 0.740 | 0.912 | 0.467 | 0.926 |
| | | 13 | 0.568 | 1.315 | -31.054 | 0.890 | 0.685 | 0.923 | 0.575 | 0.902 |
| | | 14 | 0.550 | 1.221 | -31.198 | 0.846 | 0.627 | 0.932 | 0.704 | 0.872 |
| $\lambda_B \sim$ Gamma(r,p/(1-p)) | r=28.309; p=0.377 | 27 | 0.388 | 0.633 | -23.176 | 0.882 | 0.814 | - | 0.331 | 0.847 |
| | | **28** | **0.379** | **0.611** | **-23.171** | **0.857** | **0.802** | **-** | **0.375** | **0.829** |
| | | 29 | 0.371 | 0.590 | -23.172 | 0.837 | 0.788 | - | 0.416 | 0.812 |
| | | 30 | 0.363 | 0.570 | -23.179 | 0.822 | 0.775 | - | 0.451 | 0.798 |
| | | 31 | 0.356 | 0.522 | -23.190 | 0.650 | 0.507 | - | 1.110 | 0.574 |
| | | 32 | 0.348 | 0.534 | -23.207 | 0.789 | 0.742 | - | 0.535 | 0.765 |
| | | 10 | 0.631 | 1.710 | -24.975 | 0.722 | 0.591 | - | 0.852 | 0.653 |
| | | 11 | 0.609 | 1.555 | -24.699 | 0.793 | 0.638 | - | 0.681 | 0.711 |
| | | 12 | 0.588 | 1.425 | -24.461 | 0.850 | 0.679 | - | 0.550 | 0.760 |
| | | 13 | 0.568 | 1.315 | -24.256 | 0.896 | 0.715 | - | 0.445 | 0.800 |
| | | 14 | 0.550 | 1.221 | -24.079 | 0.931 | 0.747 | - | 0.363 | 0.834 |

Table 5 give more than one alternative (for different integer values of r) for every series of data (all species, e.g. $\lambda_A$; plant species, e.g. $\lambda_P$; bacteria species, e.g. $\lambda_B$) but only one corresponds to maximum value of the likelihood (the ones in bold face). The reason is that none of them is regardless to the hypothesis of dependence, because were proofed previously that it exists a coverage distribution - the Negative Binomial distribution. In order to select the most probable values of the parameters, a similar procedure should be conducted on the Negative Binomial distribution and their results are given in Table 6.

Tab.6
Different likelihood estimates for Negative Binomial distribution parameters of species

| Hypothesis | r | p | p/(1-p) | (M)LE | $p_{K-S}$ | $p_{A-D}$ | $p_{C-S}$ | C-S | $p_{C-S}$ |
|---|---|---|---|---|---|---|---|---|---|
| Obs ~ NegBin(r,p) | 9 | 0.655 | 1.900 | -293.474 | 0.410 | 0.366 | - | 1.897 | 0.387 |
| | 10 | 0.631 | 1.710 | -293.137 | 0.461 | 0.398 | - | 1.696 | 0.428 |
| | 11 | 0.609 | 1.555 | -293.001 | 0.453 | 0.395 | - | 1.721 | 0.423 |
| | 12 | 0.588 | 1.425 | -293.008 | 0.444 | 0.372 | - | 1.801 | 0.406 |
| | 13 | 0.568 | 1.315 | -293.120 | 0.436 | 0.337 | - | 1.918 | 0.383 |
| | 14 | 0.550 | 1.221 | -293.310 | 0.428 | 0.297 | - | 2.063 | 0.357 |
| | 27 | 0.388 | 0.633 | -297.695 | 0.164 | 0.035 | - | 5.160 | 0.076 |
| | 28 | 0.379 | 0.611 | -298.034 | 0.153 | 0.030 | - | 5.384 | 0.068 |
| | 29 | 0.371 | 0.590 | -298.366 | 0.141 | 0.026 | - | 5.609 | 0.061 |
| | 30 | 0.363 | 0.570 | -298.692 | 0.131 | 0.023 | - | 5.805 | 0.055 |
| | 31 | 0.356 | 0.522 | -299.013 | 0.122 | 0.020 | - | 6.016 | 0.049 |
| | 32 | 0.348 | 0.534 | -299.322 | 0.113 | 0.018 | - | 6.198 | 0.045 |

(M): estimate of p remains the same, and thus (r,p) pair is a MLE estimate for the given r

An important remark opens a discussion here. Thus, at least one out of the two individual series - the Bacteria series - is rejected to provide reasonable likelihood estimates from its Poisson parameters (Table 6, r from 27 to 32, MLE estimate of r from $\lambda_B$ being 28). This fact excludes the opposite alternative from symmetry reasons - accepting just one alternative it means that the homogeneity hypothesis should be rejected too, which is not an acceptable result, because were proofed previously that the independence hypothesis cannot be rejected and test of independence is equivalent with test of homogeneity when Chi-Square test are involved, and it were involved. It remains that both individual series should be rejected from the simultaneous agreement Obs ~ NegBin(r,p) and $\lambda_{B\text{ or }P}$ ~ Gamma(r,p/(1/p)).

Even more, a simple calculus of the MLE estimates of p from $\lambda_B$ ~ Gamma($r_B,p_B/(1-p_B)$) and $\lambda_P$ ~ Gamma($r_P,p_P/(1-p_P)$) - values in Table 5 - gives $p_B + p_P = 0.379 + 0.631 = 1.01 \sim 1.00$ which is more than a coincidence, because the data behind $\lambda_B$ and $\lambda_P$ estimates are not independent (are the same) and thus the relationship $p_B + p_P = 1.0$ should be considered when estimates of the $r_B$ and $r_P$ are made. Consequential, the estimates from $\lambda_A$ ~ Gamma($r_A,p_A/(1-p_A)$) and Obs ~ NegBin(r,p) should be linked together, and indeed, the values of $p_A$, $r_A$ and their associated statistics from Table 5 and the values of p and r and their associated statistics from Table 6 sustain this hypothesis. The Table 7 contains the estimates using these relationships.

Tab.7
Estimates under association

| $\lambda A$ ~ Gamma($r_A,p_A/(1-p_A)$), Obs ~ NegBin($r_A,p_A$) | | | | | | | | | | | | | | |
|---|---|---|---|---|---|---|---|---|---|---|---|---|---|---|
| $\partial LE/\partial r_A=\partial LE/\partial p_A=0$ | | | | Natural r; best alternative: MLE | | | | NegBin($r_A,p_A$) | | Gamma($r_A,p_A/(1-p_A)$) | | | Global | |
| $r_A$ | $p_A$ | $p_A/(1-p_A)$ | MLE | $r_A$ | $p_A$ | $p_A/(1-p_A)$ | MLE | $p_{K-S}$ | $p_{A-D}$ | $p_{K-S}$ | $p_{A-D}$ | $p_{C-S}$ | C-S | $p_{C-S}$ |
| 12.349 | 0.581 | 1.385 | -348.399 | 12 | 0.588 | 1.425 | -348.409 | 0.467 | 0.381 | 0.999 | 0.902 | 0.974 | 1.9 | 0.869 |
| | | | | 13 | 0.568 | 1.315 | -348.430 | 0.430 | 0.334 | 0.999 | 0.909 | 0.865 | 2.2 | 0.823 |
| q=pBP/(1-pBP); $\lambda_B$ ~ Gamma($r_B,q$), $\lambda_P$ ~ Gamma($r_P,1/q$) | | | | | | | | | | | | | | |
| $\partial LE/\partial r_B=\partial LE/\partial r_P=\partial LE/\partial p_{BP}=0$ | | | | Natural r; "best": MLE | | | | Gamma($r_B,q$)) | | Gamma($r_P, 1/q$) | | | Global | |
| $r_B$ | $r_P$ | $p_{BP}$ | MLE | $r_B$ | $r_P$ | p | MLE | $p_{K-S}$ | $p_{A-D}$ | $p_{K-S}$ | $p_{A-D}$ | $p_{C-S}$ | C-S | $p_{C-S}$ |
| 29.103 | 10.030 | 0.370 | -54.000 | 29 | 10 | 0.370 | -54.001 | 0.861 | 0.790 | 0.989 | 0.832 | 0.877 | 0.71 | 0.982 |
| | | | | 30 | 10 | 0.365 | -54.029 | 0.952 | 0.798 | 0.768 | 0.764 | - | 0.81 | 0.937 |
| | | | | 29 | 11 | 0.377 | -54.322 | 0.736 | 0.580 | 0.644 | 0.707 | - | 1.64 | 0.802 |
| | | | | 30 | 11 | 0.371 | -54.599 | 0.628 | 0.447 | 0.556 | 0.634 | - | 2.31 | 0.678 |

Interpreting results given in Table 7, is no reason to reject the hypotheses that between Gamma distribution parameters of Poisson estimates of the antibacterial activities and Negative Binomial distribution of the observables it exists the relationship given by the convolution of the Poisson distribution and Gamma distribution:

$$\text{NegBin}(x;r_A,p_A) = \int_0^\infty \text{Poisson}(x;\lambda_A) \cdot \text{Gamma}(\lambda_A;r_A,\frac{p_A}{1-p_A})d\lambda_A$$

and the Gamma distribution probably occurs and characterize the interaction between these two types of organisms: plants and bacteria.

On another hand, the relationship between proportions from Gamma distribution of the Poisson parameters of the bacteria and plant series of data clearly indicate that the two factors - "bacteria factor" and "plant factor" in antibacterial activity has multiplicative and complementary effect and the separation of factors given in Table 2 has statistical sustainability. This fact opens the path to construct population factors of bacteria and plants at contingency of effects in antibacterial activity. More than that, the convolution of the two distributions, Poisson and Gamma strongly suggests that the Gamma distribution occurs due to the continuous effect of factors (as values from Table 1 are). Next table contains the parameters of the Gamma distributions of the population factors.

Tab.8
Distributions of the population factors for plants and bacteria on antibacterial activity

| Population | Distribution | r | q | q/(1-q) | MLE | $p_{K-S}$ | $p_{A-D}$ | $p_{C-S}$ | C-S | $p_{C-S}$ | $h_1[\cdot]$ |
|---|---|---|---|---|---|---|---|---|---|---|---|
| Bacteria | $a_i \sim \text{Gamma}(r_B, q_B/(1-q_B))$ | 31.663 | 0.120 | 0.137 | -10.323 | 0.816 | 0.792 | - | 0.44 | 0.804 | 1.148 |
| Plants | $b_j \sim \text{Gamma}(r_P, q_P/(1-q_P))$ | 10.082 | 0.282 | 0.392 | -16.043 | 0.993 | 0.852 | 0.898 | 0.27 | 0.965 | 1.604 |

Following figure (Figure 1) depicts the population factors distribution of plants (FP) and of bacteria (FB) and the true distribution of the antibacterial activity as convolution of these two (AA). Let's note that the convolution of two Gamma distributions only in very rare cases has a close form (expressed by a explicit distribution function) and here is not the case. It only may be approximated with another Gamma distribution.

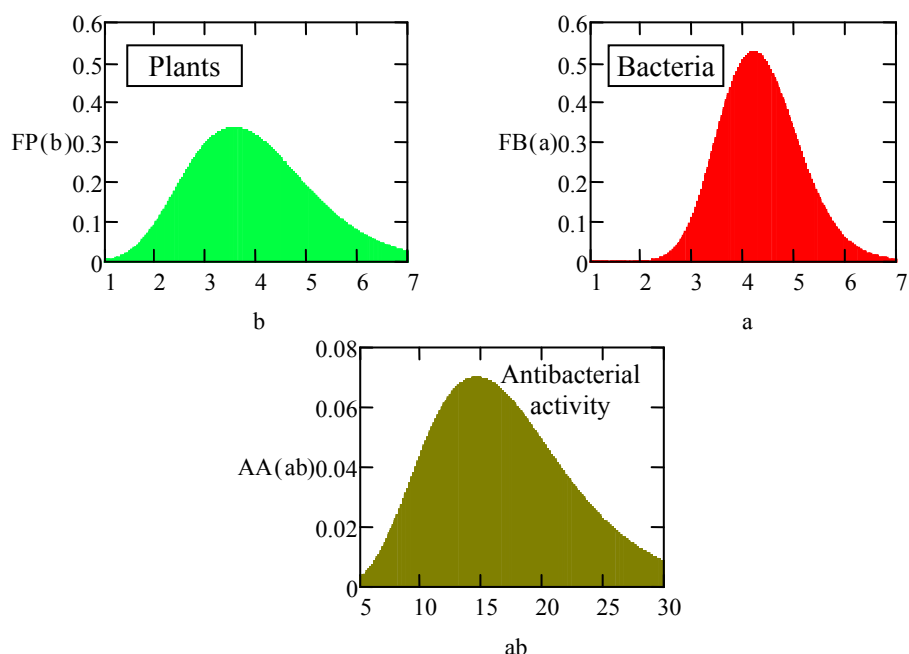

Figure 1. Population factors distribution of plants (FP) and of bacteria (FB) as well as the true distribution of the antibacterial activity as convolution of these two (AA)

As can be seen from the above figure, interesting extracted information is that the bacteria have a more slim distribution than the plants have. More, all three distributions are asymmetrical with more weight to low values (low effects, low interactions are more often between them).

**Conclusions**

The analysis of factors conducted in this study on antibacterial activity of essential oils extracts from a series of plants on a series of bacteria revealed that the Negative Binomial distribution of the antibacterial activity is a mixture (convolution) of Poisson and Gamma distributions from which only Gamma distribution can (and should) be assigned to plant and bacteria factors expressed in the antibacterial activity. Decomposition of factors under the multiplicative effect revealed a very good agreement between observed and expected values (probability of wrong model less than 0.001). Shaping of the Gamma distribution of the factors (on a relative scale) revealed that low factor values are often than high ones (left weighting of both factors distributions). The differential entropy (which is directly linked with population diversity) of plant factors are 40% higher than the differential entropy of bacteria, giving an estimate of over 50% higher diversity of plant factors than bacteria factors.

### Acknowledgments

The study was supported by POSDRU/89/1.5/S/62371 through a postdoctoral fellowship for L. Jäntschi.

### References

[1] Jäntschi L. 2009. Distribution Fitting 1. Parameters Estimation under Assumption of Agreement between Observation and Model, Bulletin UASVM Horticulture, 66(2):684-690. http://arxiv.org/abs/0907.2829

[2] Jäntschi L, Bolboacă SD. 2009. Distribution Fitting 2. Pearson-Fisher, Kolmogorov-Smirnov, Anderson-Darling, Wilks-Shapiro, Kramer-von-Misses and Jarque-Bera statistics. Bulletin UASVM Horticulture 66(2):691-697. http://arxiv.org/abs/0907.2832

[3] Bolboacă SD, Jäntschi L. 2009. Distribution Fitting 3. Analysis under Normality Assumption. Bulletin UASVM Horticulture 66(2):698-705.

[4] Jäntschi L, Bolboacă SD, Stoenoiu CE, Iancu M, Marta MM, Pică EM, Ştefu M, Sestraş AF, Duda MM, Sestraş RE, Ţigan Ş, Abrudan I, Bălan MC. 2009. Distribution Fitting 4. Benford test on a sample of observed genotypes number from running of a genetic algorithm. Bulletin UASVM Agriculture 66(1):82-88.

[5] Jäntschi L, Stoenoiu CE, Bolboacă SD, Bălan MC, Bolunduţ LC, Popescu V, Naşcu HI, Abrudan I. 2009. (Distribution Fitting 5.) Knowledge Assessment: Distribution of Answers to an Online Quizzed System. Bulletin UASVM Horticulture 66(2):680-683.

[6] Marta MM, Stoenoiu CE, Jäntschi L, Abrudan I. 2009. (Distribution Fitting 6.) Issues in Choosing a University. Bulletin UASVM Horticulture 66(2):816-822.

[7] Jäntschi L, Bălan MC. 2009. (Distribution Fitting 7.) Analysis of the genotypes number in different selection and survival strategies. Bulletin UASVM Horticulture 66(1):58-65.

[8] Bălan MC, Bolboacă SD, Jäntschi L, Sestraş RE. 2009. (Distribution Fitting 8.) Weather Monitoring: Wind Analysis (May, 2009; GPS: Lat. N46°45'35"; Long. E23°34'19"). Bulletin UASVM Horticulture 66(1):7-9.

[9] Stoenoiu CE, Bolboacă SD, Abrudan I, Jäntschi L. 2009. (Distribution Fitting 9.) Student perception of degree of academic community involvement in academic life. Actual Problems of Economics 101(11):276-283.

[10] Stoenoiu CE, Bolboacă SD, Abrudan I, Jäntschi L. 2009. (Distribution Fitting 10.) The Role of the Academic Community in Defining the Professional Route: Students' Perception. Actual Problems of Economics 103(1):277-285.

[11] Jäntschi L, Bolboacă SD. 2011. Distribution Fitting 11. Morphology of gymnosperms seeds. Manuscript send to publication.

[12] Jäntschi L, Bolboacă SD, Sestraş RE. 2011. Distribution fitting 12. Sampling distribution of compounds abundance from plant species measured by instrumentation. Application to plants metabolism classification. To appear in Bulletin UASVM Horticulture.

[13] Soković M, Marin PD, Brkić D, van Griensven LJLD. 2007. Chemical composition and antibacterial activity of essential oils of ten aromatic plants against human pathogenic bacteria. Food 1(2):220-226.

[14] Fisher RA. 1923. Studies in Crop Variation. II. The Manurial Response of Different Potato Varieties. Journal of Agricultural Science 13:311-320.

[15] Bolboacă SD, Jäntschi L, Sestraş AF, Sestraş RE, Pamfil D. Pearson-Fisher Chi-Square Statistic Revisited. To appear in Information.